\documentclass[%
 reprint,
 amsmath,amssymb,
 aps,
 pra,
]{revtex4-2}

\usepackage{graphicx}
\usepackage{dcolumn}
\usepackage{bm}
\usepackage{siunitx}
\usepackage{amsmath}
\usepackage[english]{babel}
\usepackage{titlesec}
\usepackage{graphicx}
\usepackage{dcolumn}
\usepackage{bm}

\usepackage{siunitx}
\begin{document}
\preprint{APS/123-QED}

\title{CMOS-Fabricated Ultraviolet Light Modulator\\ Using Low-Loss Alumina Piezo-Optomechanical Photonics}

\author{Zachary A. Castillo$^{1,2\ddag}$}
\author{Roman Shugayev$^{1\ddag}$}
\altaffiliation[Now at ]{Department of ECE, University of Nevada, Las Vegas}
\author{Daniel Dominguez$^1$}
\author{Michael Gehl$^1$}
\author{Nicholas Karl$^1$} 
\author{Andrew Leenheer$^1$}
\author{Bethany Little$^1$}
\author{Yuan-Yu Jau$^1$}
\author{Matt Eichenfield$^{1,3}$}\email{eichenfield@arizona.edu}

\affiliation{$^1$Microsystems Engineering, Science, and Applications (MESA), Sandia National Laboratories, Albuquerque, NM 87185, USA \\ $^2$ Department of Physics and Astronomy, Center for Quantum Information and Control, University of New Mexico, Albuquerque, NM 87131, USA\\$^3$Wyant College of Optical Sciences, University of Arizona, Tuscon, AZ, 85721, USA\\$^\ddag$ These authors contributed equally to this work.}

\begin{abstract}
Ultra-violet (UV) and near-UV wavelengths are necessary for many important optical transitions for quantum technologies and various sensing mechanisms for biological and chemical detection. However, all well-known photonic platforms have excessively high losses in the UV, which has prevented photonic integrated circuits (PICs) being used to address these and other important application spaces. Photonic waveguides using low-loss alumina cores have emerged as a promising solution because of almunia's large optical bandgap and the high quality of films enabled by atomic layer deposition. However, to the best of our knowledge, \textit{active} alumina PICs have only been realized using thermo-optic tuning, which precludes switching speeds shorter than one microsecond, high circuit densities, and cryogenically compatible operation. Here, we introduce a CMOS-fabricated, piezo-optomechanical PIC platform using alumina waveguides with low optical losses at UV wavelengths and aluminum nitride piezoelectric strain actuators, which solves the issues associated with thermal tuning. We demonstrate a high-performance, reconfigurable optical filter operating at wavelengths as low as 320 nm. The filter has a 6 nanosecond switching time, a loaded linewidth of 3.3 GHz, tuning rate of -120 MHz/V, and a hold power less than 20 nW.  \end{abstract}

\maketitle

Applications that require near-ultraviolet (near-UV) and UV light include important optical transitions for quantum computing \cite{q_compute}, sensing \cite{q_sens}, and timekeeping \cite{q_clock}; fluorescence-based assays for protein \cite{assay_bio} and chemical \cite{assay_chemical} identification; and UV enhanced Raman spectroscopy \cite{raman}. Despite the numerous applications which can benefit from the dramatically reduced size, increased stability, and potential for mass manufacture that photonic integrated circuits (PICs) provide compared to fiber and bulk optical implementations, demonstrations of UV photonic circuits have been few and far between due to limited selection of waveguiding materials and stringent fabrication demands. Alumina films made by atomic layer deposition are standout materials for UV photonics because of CMOS-compatible processing, single dB/cm losses down to 225 \unit{\nano\meter} \cite{alumina_2010}, and the ability to fully \cite{Alumina_2019,Alumina_high_confinement} or partially \cite{Alumina_high_Q} etch waveguides while maintaining smooth sidewalls. Active alumina-based PICs using thermo-optic tuners have been demonstrated for UV microscopy \cite{UV_SIM} and UV laser narrowing and tuning \cite{UV_Laser}. However, thermo-optical tuning is often undesirable due to thermal crosstalk between components that prevents high circuit density and relatively slow thermal time constants that exclude applications that require response times faster than 1 microsecond \cite{thermo2}. Additionally, PICs using heaters are not suitable for cryogenic applications, as the dissipated heat can overwhelm cooling budgets, and the thermo-optic coefficients of waveguide materials decrease exponentially at low temperatures \cite{thermo_cryo}, requiring large temperature excursions and/or long waveguide lengths.
\begin{figure*}
    \centering
    \includegraphics[scale=1]{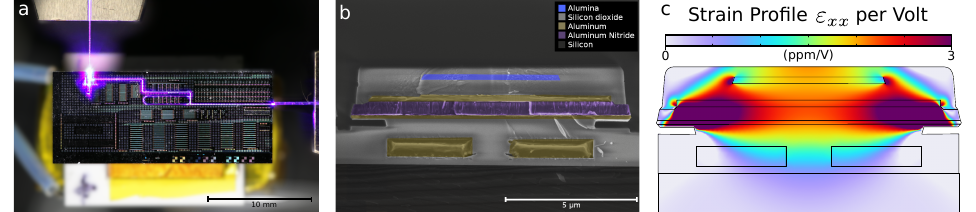}
    \caption{{\bf A piezo-optomechanical racetrack resonator.} {\bf a} Photograph of a 21 x 8 mm photonic test chip with in-coupled 420 nm light. {\bf b} False colored SEM cross section of the piezo actuated portion of the resonator showing the relevant material layers. {\bf c} Finite element simulation result showing the distribution of the of the dominate strain tensor component per volt for DC actuation.}
    \label{fig:pic}
\end{figure*}
Here we demonstrate a fully CMOS-fabricated platform for piezo-optomechanical UV PICs, capable of low losses at UV wavelengths, switching times faster than 10 nanoseconds, and orders of magnitude smaller power consumption than thermal modulation. Waveguides using fully etched alumina thin film cores deposited by atomic layer deposition are strain tuned by an integrated aluminum nitride piezoelectric actuator, which allows high-speed and low-power tuning. As a demonstration of a UV compatible piezo-optomechanical PIC, we fabricated and characterized a reconfigurable photonic notch filter based on a strain-tuned racetrack resonator. At 320 nm, the resonator shows low propagation losses (at most, 2.5 dB/cm, corresponding to a 3.3 GHz loaded linewidth), a resonance tuning rate of -120 MHz/V, and a switching time of 6 \unit{\nano\second}. This is, to the best of our knowledge, the first demonstration of a piezo-optomechanical PIC compatible with UV wavelengths, and a high-performance reconfigurable PIC filter with potential to operate at wavelengths as low as 225 nm \cite{alumina_2010}. 

Notably, piezoelectric strain actuation of photonic waveguides has been demonstrated with silicon nitride \cite{bowers,stanfield} and silicon \cite{yale_silicon} waveguides, allowing demonstrations of tunable filters \cite{bowers}, optically broadband phase and amplitude modulators \cite{cMZI}, and compact acousto-optic modulators (AOMs) \cite{yale_silicon,jake}. These basic PIC elements have been further integrated into larger circuits for important applications including programmable interferometer meshes \cite{cryo},  strain and optical control of vacancy centers \cite{spin_phonon, artificial_atom}, photonic modulators for trapped ions \cite{ion}, photonic isolators \cite{isolator}, actuation of soliton frequency combs \cite{comb}, and many others \cite{review} . It is reasonable to expect that similar functions can be implemented using alumina waveguides, providing a path forward for a diverse number of previously inaccessible applications that can benefit from rapidly tunable PICs at near-UV/UV wavelengths.

Our fabricated piezo-optomechanical filter is shown in Fig. \ref{fig:pic}a. while being interrogated with 420 nm light coupled in via lensed fiber. The filter is a tunable racetrack resonator consisting of two straight, piezo-optomechanically tunable sections with 5 \unit{\micro\meter} wide multimode (MM) cores, 3.45 \unit{\milli\meter} in length. The resonator bends were semicircular arcs with single-mode (SM) 250 nm wide cores and 400 \unit{\micro\meter} radii of curvature. SM bends were used to maintain high coupling ideality with the bus waveguide \cite{coupling}. The SM to MM transitions were 200 \unit{\micro\meter} long quadratic tapers.

\begin{figure*}
  \centering
  \includegraphics[scale=1]{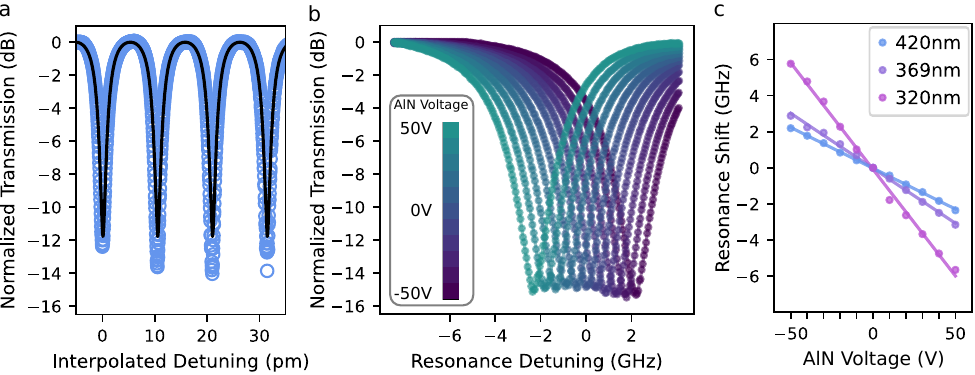}
  \caption{{\bf Transmission and static piezo-tuning.} {\bf a} Normalized transmission of the filter near 420 nm acquired by thermal tuning. The propagation loss is extracted from the fit (black line). {\bf b} Voltage tuning near 420 nm. {\bf c} Consolidated resonance shifts with linear fits.}
  \label{fig:data}
\end{figure*}

The key elements of the piezo-optomechanical structure are highlighted in the false-color scanning electron micrograph (SEM), Fig. \ref{fig:data}b. The waveguides were fabricated with 150 nm thick alumina cores deposited by atomic layer deposition at 300\unit{\degreeCelsius}, surrounded by silicon dioxide cladding. Detailed dimensions are provided in the supplement. Crucial to the platform is the piezoelectric transducer, formed by a polycrystalline but highly oriented aluminum nitride (AlN) layer with top and bottom aluminum electrodes. Integrated vertical electrical routing using a combination of tungsten and aluminum vias allows independent electrical connections between a low resistivity routing layer to either electrode with sub-micron footprints. The devices were undercut to increase the mechanical compliance by isotropic etching a buried amorphous silicon layer with $\text{XeF}_2$. Based on simulations using finite element modeling software (FEM), we estimated that we can generate strains of approximately $2\times10^{-6}$/\unit{\volt} near the center of the waveguide (Fig. \ref{fig:pic}c.). The devices were fabricated in Sandia National Laboratories' 200 mm MESA volume-CMOS. 

Our primary goal was to characterize the optical linewidth, tuning rate, modulation bandwidth, and switching time of the filter at 320 nm, the wavelength the device design was optimized for. Our laser source was not tunable, thus we utilized an indirect approach to measure the linewidth and tuning rate. To this end, we thermally tuned the temperature of the photonic chip using closed feedback to an external Peltier element across several free spectral ranges (FSRs). Using the fact that the FSR is nearly constant over our thermal sweep, we calibrated the chip's temperature to an effective detuning using a numerically calculated FSR from FEM. Because the FSR ($\approx5 \text{ pm}$) is not much larger than the linewidth of the filter ($\approx1 \text{ pm}$), the spectral response is not Lorentzian. Therefore, we fitted the transmittance data, $T$, to a more general model, parameterized by a cross-coupling coefficient, $t_1$, the round trip transmission ratio, $t_2$, and the round trip optical phase, $\phi$ (Eq. \ref{model_fit_main})\cite{fit_model}.
\begin{equation}\label{model_fit_main}
    T =  \left(\frac{t_1^2+t_2^2-2(t_1t_2)\cos{(\phi)}}{1+(t_1t_2)^2\cos{(\phi)}}\right)
\end{equation}
Using the fitted parameters, $t_1$ and $t_2$ we calculated the loaded linewidth of 3.3 GHz (loaded $Q=280,000$). The coupling condition for our tested device was ambiguous, so we were unable to distinguish $t_1$ from $t_2$. Using the smaller of the two values, the intrinsic loss is at most $4.4\pm0.4$\unit{\decibel}/\unit{\centi\meter}, corresponding to an intrinsic quality factor of at least $Q=350,000$. 

As a post facto check, we repeated the experiment and fitting procedure near 369 and 420 nm. The lasers used at these wavelengths had sufficient tuning to directly scan the wavelengths across a full FSR, allowing us to compare our calculated FSRs and fitted linewidths to direct measurements. We found agreement within 10\unit{\percent} at these wavelengths, with the fitted parameters giving slightly overestimated values. Furthermore, we observe similarly low intrinsic losses at the single dB/cm level. Representative data taken near 420 nm using the indirect approach is shown in Fig. \ref{fig:data}a. Further details about our methods and the intrinsic losses are discussed in Supplement 1.

Next, we characterized the static piezoelectric tuning response of the device by applying DC voltage to strain tuning electrodes (Fig. \ref{fig:data}b). We performed linear fits of the shifts in the transmission minima against applied voltage to calculate the tuning rates (Fig. \ref{fig:data}c.), yielding -120 MHz/V at 320. We then measured the leakage resistance of the actuators using a source measure unit sourcing 20 \unit\volt, giving a value of at least 52 \unit{\giga\ohm} at room temperature. Assuming a constant leakage resistance for slightly higher voltages, we expect a hold power of approximately 15 nW or less to keep the resonator detuned by one linewidth at 320 nm.

\begin{figure*}
  \centering
  \includegraphics[scale=1]{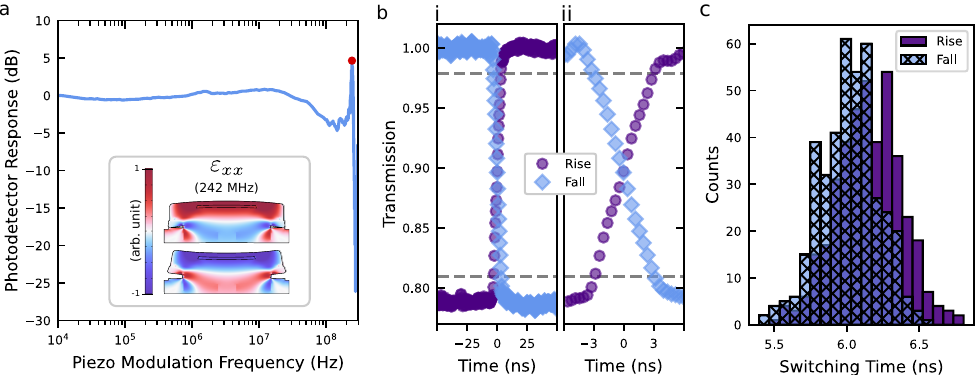}
  \caption{{\bf Electrical bandwidth and switching time.} {\bf a} Normalized intensity modulation response with an especially strong response at 242 MHz (red data point). {\bf inset} Simulation result showing the relevant strain (colored) and exaggerated displacement (edge) profiles of the strongly coupled mechanical eigenmode responsible for the red data point. {\bf b} Representative rise (purple circles) and fall (blue diamond) time data over a 100 ns ({\bf i}), and a 12 ns interval ({\bf ii}). The dashed line indicates the threshold values used for the measurement. {\bf c} Histogram of 400 measurements, showing a rise and fall time of $6.15 \pm 0.22$ and $6.02 \pm 0.20$ ns respectively.}
  \label{fig:data_switch}
\end{figure*}

Finally, we measured the modulation bandwidth and switching time of the filter. Using an electronic spectrum analyzer with an internal tracking generator, we applied a constant-amplitude sinusoidal voltage to the strain tuning electrodes and monitored the optical intensity fluctuations at the steepest point of the transmission profile. The normalized response is plotted in Fig. \ref{fig:data_switch}a. The actuation is approximately flat (within 1 dB of the baseline) until approximately 60 MHz, at which point a roll-off consistent with the actuator's RC time constant from metal trace resistance and capacitance becomes significant. The roll-off is interrupted by resonantly enhanced mechanical eigenmodes of the structure near 100 MHz, with an especially strongly coupled mode at 242 MHz. Because the modulation response is not a simple one-pole roll-off, it is important to directly measure the switching time of the filter. With this in mind, we applied a 10V peak-to-peak square wave voltage (3 ns rise/fall time) and monitored the transmission change with a high bandwidth oscilloscope. Representative rising and falling edges are shown in Fig. \ref{fig:data_switch}b. We quantified the switching time by measuring the time between the 10\% and 90\% threshold values, calculated with respect to the change in steady state transmission levels, and find an average rise and fall time value of approximately 6 ns (Fig. \ref{fig:data_switch}c.).
To better understand the device performance, we created multiphysics models using FEM and calculated the expected resonance tuning rates. The fractional shift in the resonance frequency of a resonator under the influence of piezoelectric strain can be calculated using first-order perturbation theory \cite{moving_boundary}, and can be expressed as
\begin{equation}
    \frac{\Delta\omega}{\omega_0} =-\frac{1}{2}\frac{\int dV \left(\Delta\epsilon(\vec{r}) |\vec{E}(\vec{r})|^2 \right)}{\int dV \left(\epsilon(\vec{r}) |\vec{E}(\vec{r})|^2 \right)}
\end{equation}
where $\omega$ is the angular frequency of the mode, $\vec{E}$ is the position dependent electric field of the resonator mode, $\epsilon(\vec{r})$ is the position dependent relative permittivity tensor, and $\Delta\epsilon(\vec{r})$ quantifies the strain induced perturbations to the waveguide. Although one could, in principle, create a piezoelectric-mechanical-electromagnetic coupled FEM model, solve for the piezoelectrically induced strain and resultant dielectric changes everywhere, and integrate the expression above, this is computationally intractable in our case due to the large size of the resonator relative to the wavelength (roughly 40,000 wavelengths at 320 nm). 
Instead, we approximated the calculation by considering the perturbation to the propagation constant in the piezoelectrically actuated straight sections, where the strain is uniform along the direction of propagation, allowing us to reduce the problem to calculating the change in phase accumulated over those lengths and the resultant change in the resonator frequency. To wit, 
\begin{equation}\label{perturbation_beta}
    \frac{\Delta \omega}{\omega} = -\frac{2L\Delta\beta }{2\pi R \beta^{(0)}_{\text{bend}}+2L\beta^{(0)}_{\text{straight}}+4\int ds (\beta^{(0)}(s))}
\end{equation}
where $L$ is the length of the actuated portion, $\Delta\beta$ is the perturbation to position dependent  angular wave number $\beta^{(0)}$, $R$ is the radius of the waveguide arcs, $\beta^{(0)}_{\text{bend}(\text{straight})}$ is the effective mode index of the curved (straight) portions, and the integral in the denominator is taken along the waveguide taper connecting the SM and MM waveguides. The change in wavenumber along the actuated sections has two contributions, the first being the photoelastic effect ($\Delta \beta_{PE}$) \cite{photo_elastic} that accounts for the stress-dependent index of refection in materials, and the second being the moving boundary effect ($\Delta \beta_{MB}$) \cite{moving_boundary} which accounts for geometric changes. Our model indicates an expected tuning rate of -350 MHz/V at 320 nm, which is 2.9 times higher than what we observed. Possible reasons for this discrepancy, further details about the simulation model, and the and material values used for the simulation are discussed in Supplement 1.

The device discussed here shows promise as a useful amplitude modulator in integrated photonic circuits operating at UV wavelengths because of volume-CMOS fabrication, orders of magnitude smaller power dissipation compared to thermo-optic tuning, and switching times relevant for applications such as qubit preparation and readout. Still, there is room for improvement. The extinction ratio was at most 7 dB near 320 nm, but a redesign of the coupler paired with knowledge of typical losses in the platform will readily improve this. In addition, it is advantageous to reduce the necessary driving voltages. This can be addressed by improving the resonator linewidth and increasing the piezo-responsivity. Assuming that a propagation loss of 0.1 dB/cm is achievable at 320 nm, we expect a loaded linewidth of $~200 \text{ MHz } (Q \approx  4 \text{ million})$ which would allow approximately one linewidth per volt tuning based on the current tuning rate. Further reductions in drive voltage could be achieved by larger structural undercuts, or by using scandium-doped AlN, which has shown a five-fold improvement in the $d_{33}$ coefficient compared to undoped AlN \cite{AlScN}. Despite these shortcomings, the modulator performs comparably to commercially available free-space UV-compatible AOMs or Pockels cells in terms of switching speeds, while eliminating the need for free space beams, and the watt-level powers or kV-level voltages typically required by these devices, respectively.

Looking further, the low loss of the alumina waveguides down to 225 nm \cite{alumina_2010} coupled with the low-power active tuning provided by the piezo layer allows a multitude of useful near-UV/UV photonic elements and circuits. Examples include other types of low-power and fast modulator PICs including optically broadband phase and amplitude modulators, polarization modulators \cite{polarization}, and gigahertz frequency AOMs \cite{jake}. These devices can provide the basic units of reconfigurable photonic circuits for quantum technologies using neutral atoms or ions with near-UV/UV optical transitions. Applications beyond quantum sciences include programmable photonic circuits for lab-on-a chip or waveguide-enhanced Raman spectroscopy systems \cite{raman_wg}, non-line-of-sight communication/imaging \cite{non-line-of-sight}, beam profile shaping using optical phased arrays \cite{bessel}, and off-chip beam scanning structures \cite{beam_scan}. 

\vspace{1em}
See Supplement 1 for supporting content.

\begin{acknowledgments}
The authors thank Matthew Chow for experimental discussion, Matthew Koppa for assistance with device modeling, and Sam Peana for assistance with data analysis. 
This work was performed, in part, at the Center for Integrated Nanotechnologies, an Office of Science User Facility operated for the U.S. Department of Energy (DOE) Office of Science. This material is based upon work supported by the U.S. DOE, Office of Science, National QIS Research Centers, Quantum Systems Accelerator,  and by the Laboratory Directed Research and Development program at Sandia National Laboratories (SNL). SNL is a multiprogram laboratory managed and operated by National Technology and Engineering Solutions of Sandia, LLC., a wholly owned subsidiary of Honeywell International, Inc., for the U.S. DOE’s National Nuclear Security Administration under contract DE-NA-0003525. This paper describes objective technical results and analysis. Any subjective views or opinions that might be expressed in the paper do not necessarily represent the views of the U.S. DOE or the U.S. Government.
\end{acknowledgments}
\bibliographystyle{ieeetr}
\bibliography{main_article}
\end{document}

\begin{figure*}[ht]
    \centering
    \includegraphics[scale=1]{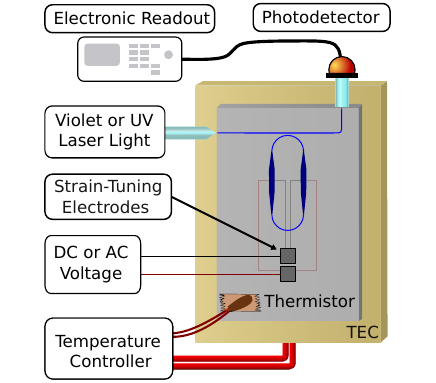}
    \caption{{\bf Testing Schematic.}}
    \label{fig:test}
\end{figure*}

\begin{center}
    {\bf \Large CMOS-Fabricated Ultraviolet Light Modulator Using Low-Loss Alumina Piezo-Optomechanical Photonics: Supplement 1}
\end{center}
\begin{center} {\bf \large S1: Device Characterization} \end{center}
The testing setup is shown schematically in Fig. \ref{fig:test} and described in detail in the following subsections. 

\begin{center} {\bf S1-A: Fiber Coupling and Bus Waveguide} \end{center}
Light was launched into a 500 \unit{\nano\meter} wide waveguide core using a polished lensed fiber (S405-XP) with a focal spot diameter of approximately 2 \unit{\micro\meter}. The core width at the input facet linearly tapered to 250 nm over 600 \unit{\micro\meter}. This 250 nm width was kept constant over 2 cm which included the coupling region with the resonator. The coupling gap was drawn to be 205 nm between the interior edges of the two waveguide cores at the apex of the resonator apex. The bus waveguide core tapered back to a 500 nm width at the output facet, over an 8 mm length. The input and output facets of the bus waveguide were designed at a right angle to minimize the amount of stray light collected. We collected the output light using a large NA multimode fiber and connected the fiber to a UV-enhanced silicon photodetector.  We observed approximately 14 dB of total insertion loss at the ouptut facet when 420 nm laser light was tuned to maximize the transmittance. Further measurements are needed to quantify the amount of loss attributable to insertion losses from optical coupling and from propagation loss in the passive single-mode waveguides. This will be investigated in future work.

\begin{center} {\bf S1-B: Electronic Readout} \end{center}
A fiber coupled, UV-enhanced avalanche photodetector with DC-400 MHz bandwidth was connected to a digital acquisition card (DAQ) for the thermally tuned optical transmission and the DC tuning measurements. The maximum sample rate of the DAQ card was inadequate for the high-speed modulation and switching time measurements. Instead, the detector was connected to the input of a swept source electronic spectrum analyzer for the electrical modulation measurements, and a high-speed (1 GHz analog bandwidth) oscilloscope for the switching time measurements.

\begin{center} {\bf S1-B: Voltage Actuation} \end{center}
We actuated the piezoelectric layer using radio frequency ground-signal-ground-signal-ground probes that made contact with square electrode pads with 100 \unit{\micro\meter} side lengths. The electrode pads were fabricated with integrated tungsten vias that were connected to a low-resistance aluminum routing layer. Electrical routing traces were patterned to connect the routing layer to the top or bottom piezo electrodes with additional vias.

We used three distinct pieces of electronic test equipment to actuate the device, depending on the measurement. A precision DC voltage source was used for the DC tuning measurements. The output port of a swept source electronic spectrum analyzer was used for the measurement of electrical modulation bandwidth. We used an arbitrary waveform generator to generate square waves with a 10 \unit{\volt} peak-to-peak amplitude and 1 \unit{\mega\hertz} frequency for the switching time measurements. We confirmed that the rise/fall time of the driving waveform was faster ($\approx3$ \unit{\nano\second}) than the observed device response ($\approx6$ \unit{\nano\second}). The 3 \unit{\nano\second} rise/fall time was observed when the function generator was connected to both 50 \unit{\ohm} and high-Z loads.

\begin{center} {\bf S1-C: Temperature Control} \end{center}

The photonic test chip (Fig. \ref{fig:test}, gray box) was secured to a Peltier module (yellow box) using double-sided polyimide tape. A 10 \unit{\kilo\ohm} negative temperature coefficient thermistor was attached to the surface of the chip using polyimide tape, and was connected to a commercial temperature controller to provide closed-loop feedback of the chip's temperature. The temperature was kept constant with stability better than 1 \unit{\milli\kelvin} for voltage tuning, modulation, and switching measurements. For the thermal tuning measurements, we changed the temperature of the chip at a rate of approximately 12 \unit{\milli\kelvin/\second}. This rate was slow enough so that the temperature difference between the resonator and the thermistor was always negligible, ensuring repeatable temperature vs transmission sweeps.

\pagebreak
\begin{center} {\bf S1-D: Laser Systems} \end{center}

\begin{figure*}[ht]
    \centering
    \includegraphics[scale=1]{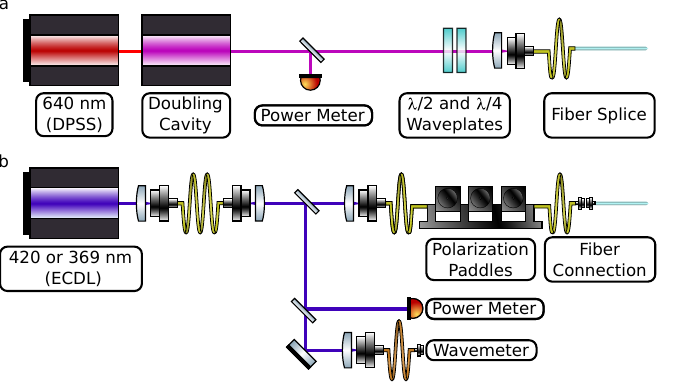}
    \caption{{\bf Laser System Schematics.} {\bf a} Schematic for 320 nm light generator. Note the free space wave plates used for polarization controls, and the fiber spice. {\bf b} Schematic for ECDL lasers. We collimate the laser sources and use the reflections from glass microscope slides to generate reference beams for wavelength and power monitoring.}
    \label{fig:laser}
\end{figure*}

The laser systems are shown schematically in Fig. \ref{fig:laser}. Laser light at 320 nm was generated by frequency doubling a high-power diode-pumped solid-state (DPSS) laser using a commercial doubling cavity, resulting in up to 100 mW of power when well aligned. A reflection from a thin glass microscope slide was used for power monitoring. Free space quarter- and half-wave plates (zero-order at 308 nm) were used to set the polarization. We fiber coupled the 320 nm light into an end-capped single-mode fiber (SM300). The end-capped fiber was necessary to prevent damage at the bare fiber end face from the intensely focused beam needed to achieve acceptable ($\approx$40\%) fiber coupling efficiency. The SM300 fiber was fusion spliced to a lensed fiber in order to minimize insertion losses from butt coupling.

Laser light at 369 and 420 nm was generated from commercial extended cavity diode lasers (ECDLs). Because of the lack of commercially available fused fiber coupler at violet and near-UV wavelengths, we collimated the fiber coupled laser outputs and generated reference beams for power and wavelength monitoring using the reflections from thin microscope slides. An external commercial wavemeter allowed us to directly measure cavity linewidths and free spectral ranges (FSRs) while we slowly scanned the voltage-controlled ECDL. We refocused most of the optical power back into a single-mode fiber. This fiber was wound through fiber polarization paddles for polarization control. The fiber was then butt-coupled to a lensed fiber for chip coupling.


\begin{center} {\bf \large S2: Resonator Transmission Model and Loss Measurements} \end{center}\label{loss}
As mentioned in the main text, we calculated the reported linewidth and intrinsic loss at 320 nm using parameters found by fitting thermally tuned transmission data. In this section, we describe the transmission model, the optical power normalization procedure, and the transformation from test chip temperature to effective cavity detuning. 

Our analysis required knowledge of the FSR of the cavity, which we estimated at 320 nm using   finite element modeling software (FEM). As a post facto check, we repeated the thermally tuned transmission measurements and fitting procedure at 369 and 420 nm. Although we directly measured the FSRs at these wavelengths, we used simulated FSRs at all three wavelengths when interpolating the temperature of the chip into an effective de-tuning for consistency. We discuss how these simulated values were acquired in Section S3-B. The directly measured FSRs are compared with the simulated FSRs in Section S2-E and Table \ref{tab:FSR}.

\begin{center} {\bf S2-A: Model} \end{center}
 The general form of the model\cite{fit_model} we used describes the transmitted power past the cavity is given as 
\begin{equation}
T = \left| \frac{E_{out}}{E_{in}} \right|^2 = \left|\frac{t_1-t_2\exp{(-i\phi)}}{1-t_1t_2\exp{(i\phi)}}\right|^2\label{model_exp}
\end{equation}
where $T$ is the optical power transmittance, $|E_{out(in)}|^2$ is proportional to the optical power exiting (entering) the resonator, $t_1$ is the electric field cross-coupling coefficient between the resonator and bus waveguide, $t_2$ is the electric field transmission coefficient after one round trip of the resonator, and $\phi$ is the round-trip optical phase delay. We assumed, without loss of generality, that $t_1$ and $t_2$ are real numbers for simplicity. We multiplied this standard expression by a prefactor so that the transmission value always maximizes to one, which was necessary to fit the normalized data (Eq. \ref{eq:norm}). 

\begin{equation}\label{eq:norm}
 T_{\text{normalized}} =  \left(\frac{1-t_1t_2}{t_1-t_2}\right)^2\left|\frac{t_1-t_2\exp{(-i\phi)}}{1-t_1t_2\exp{(i\phi)}}\right|^2 
\end{equation}
We also have the relationship
\begin{equation}
    \phi(\Delta \lambda)=\left(\frac{2\pi n_{eff}}{\lambda_0+\Delta \lambda}\right) L\label{phi}
\end{equation} 
where $n_{eff}$ is the position dependent effective refractive index, $\lambda_0$ is the laser test wavelength in vacuum, $\Delta\lambda$ is the wavelength detuning, and $L=1.02$\unit{cm} is the geometric path length. Strictly speaking, $\phi$ is changed through thermo-optic changes ( $n_{eff}\rightarrow n_{eff}+\Delta n_{TO}$) and not changes in the laser wavelength for the thermally swept transmission measurements. However, we can consider this thermo-optic shift to be equal to an effective detuning by the relationship $\Delta\lambda = -\lambda(\Delta n_{TO}/n)$ in the limit of small index changes. 

The field transmission parameter $t_2$ is related to the absorption coefficient of the waveguide, $\alpha$ according to Eq. \ref{loss_formula}.
\begin{equation}\label{loss_formula}
    t_2^2 = e^{-\alpha L}
\end{equation}
Solving for the absorption coefficient, we have
\begin{equation}\label{eq:absorption}
    \alpha = -\frac{2\ln{\left(t_2\right)}}{L}
\end{equation}
or, in terms of power loss
\begin{equation}\label{dB/cm}
    \alpha_{dB/cm} = \frac{10}{\ln(10)}\alpha =\frac{20}{\ln{(10)}}\cdot\frac{\ln{\left(t_2\right)}}{L}
\end{equation}
where $L$ is given in centimeters in Eq. \ref{dB/cm}.

Finally, we can relate Eq. \ref{eq:absorption} to the intrinsic quality factor,$Q_i$, with Eq. \ref{eq:Q},
\begin{equation}\label{eq:Q}
    Q_i = \frac{2\pi \bar{n}_g }{\alpha\lambda_0}    
\end{equation}
where $\bar{n}_g$ is the averaged group index \cite{intrinsic_Q}. We refer to an averaged group index since the group index changes along the resonator because of the varying core width. The averaged group index is simply the group index which is consistent with the free spectral range, $\nu$,
\begin{equation}
    \bar{n}_g = \frac{c}{\nu L}
\end{equation}
where $c$ is the speed of light.

\begin{figure*}[ht]
    \centering
    \includegraphics[width=\textwidth]{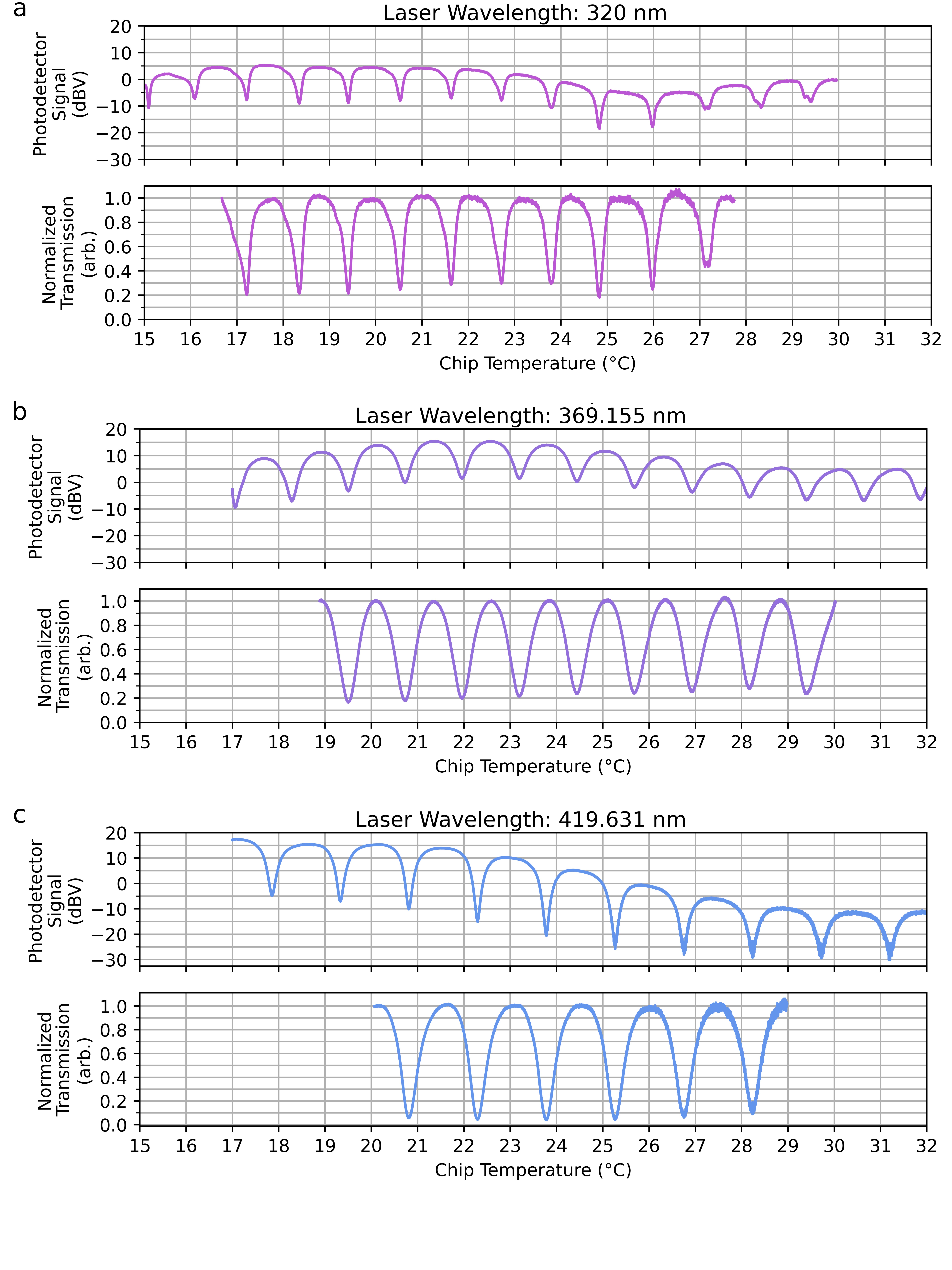}
    \caption{Raw and normalized transmission data from thermal tuning of the PIC chip.}
    \label{fig:data_conditioning}
\end{figure*}

\begin{center} {\bf S2-B: Transmitted Power Normalization} \end{center}
The raw and normalized thermally tuned transmission spectra are presented in Fig. \ref{fig:data_conditioning}. The raw data showed noticeable background deviations because of slight drift between the lensed fiber and the photonic chip as the temperature changed. To correct for this, we identified the number of resonances, $N$, present in the data set by counting the number of dips. We then fit an $N-1$ degree polynomial to the set of middle points between each pair of adjacent resonances.  Dividing the data set by the polynomial fit normalized the data so that the transmission reached one between each resonance. Overshoot in the polynomial fit at the start and end of the data set causes noticeable skewness in the transformed resonances, so we discarded conditioned data at the beginning and end of the data set. Specifically, we discarded data up to the second local transmission maxima, and data occurring after the second to last transmission maxima. We refer to this transformed data as the conditioned data in the following section.

\begin{center} {\bf S2-C: Interpolation Between Temperature to Laser Detuning} \end{center}
We converted the measured temperature of the device to an effective cavity/wavelength detuning. To find the mapping between temperature and effective detuning, we arbitrarily set the minima of the first conditioned resonance to zero detuning. We then used the calculated values of the FSR to assign each transmission minima a detuning, knowing that each dip is separated by the FSR. Finally, we interpolated the temperature values between transmission minima into effective detunings using cubic splines. We assume a constant FSR near each wavelength which is justified since the total detuning (less than 100 pm) is much smaller than the laser wavelengths. Obviously, this approach requires an accurate value of the FSR. In order to judge the accuracy of the simulated FSR values, we directly measured the FSRs at 369 nm and 420 nm using the ECDL lasers and independent wavemeter. The differences between the simulated and measured values are discussed in Section S2-E. We found a relative difference of $3.5\%$ and $5.7\%$  with respect to the measured values at 369 nm and 420 nm, respectively, giving us confidence in our simulated FSR at 320 nm.

\newpage
\begin{center} {\bf S2-D: Intrinsic Losses} \end{center}
It is worth stressing that one cannot distinguish between the over-coupled ($t_2>t_1$) and under-coupled ($t_1>t_2$) regimes using only Eq. \ref{model_exp}. Re-expressing in terms of purely real numbers, 
\begin{equation}
    T =  \left(\frac{t_1^2+t_2^2-2(t_1t_2)\cos{(\phi)}}{1+(t_1t_2)^2\cos{(\phi)}}\right)\label{model_real}
\end{equation}
we see that the transmission value is the same when $t_1$ and $t_2$ are interchanged. Therefore, there is an ambiguity in which parameter corresponds to which physical process (round trip transmission vs cross-coupling) without additional information. To put this another way, suppose that the fitting procedure returns the parameters $t'$ and $t''$. If we choose $t'$ to represent the cross-coupling ratio, then $t''$ represents the round trip transmission, and is related to the intrinsic quality factor, $Q_i$, of the resonator. However, we could simply interchange $t'$ for $t''$ and have the same transmission, meaning $Q_i$ is now related to $t'$, which is inconsistent.

A standard solution to this problem would be to test multiple nominally identical devices, but with varying coupling gaps. However, because of the large foot print of the device, and the limited space on the reticle, this could not be done. There are experimental methods to distinguish between the over- and under-coupled cases without prior knowledge by examining the optical phase near resonance. For example, one can rapidly phase modulate the input light and look at the transient response \cite{coupling_phase_mod}. This would require UV phase modulators capable of switching with speeds faster than $2\pi Q/\omega_0 \approx100,000/940 \text{ THz}=400  \text{ ps}$, but such modulators are not readily available at these wavelengths. Alternatively, one can simulate the cross-coupling coefficient using accurate knowledge of the device geometry and finite element modeling (FEM) or finite difference time domain solvers. We pursued the simulation approach, but we found that inferring the fabricated geometry with the precision necessary for accurate simulation results is intractable. See Sections S3-D and S3-E for details.

Still, we can be confident that the intrinsic loss is on the order of a few dB/cm at our tested wavelengths. Table \ref{tab:model_fits} lists the extracted model parameters for the fitted thermally swept data. We refer to the parameters $t'$ and $t''$ instead of $t_1$ and $t_2$ because of the ambiguity discussed above. Using both values, we calculated the possible intrinsic propagation losses and intrinsic quality factors. We are inclined to believe that the intrinsic losses are 4.4, 4.3, and 3.3 dB/cm at 320, 369, and 420 nm respectively, corresponding to $t'$ at 320 nm, and $t''$ at 369 and 420 nm. This is supported by the logical trend of decreasing loss as a function of wavelength, and comparable intrnsic quality factors. As points of comparison, a loss of 3.2 dB/cm was reported in \cite{Alumina_2019} at 371 nm for a fully etched single mode waveguide and, more recently, a remarkably low loss of 0.6 dB/cm at 360 nm was reported in \cite{alumina_CMOS}, again for fully etched single mode waveguides. Reference \cite{Alumina_high_confinement} reported a loss of 0.8 dB/cm for a fully etched, but multimode waveguide, at 450 nm for the fundamental TE mode. 

Finally, we estimated the loss contribution from the top metal electrode. We performed a two-dimensional wavelength-dependent mode analysis of the piezo-actuated cross section of the resonator (Fig. 1c, main text), including the top electrode in the simulation domain. We plot the calculated propagation loss as a function of wavelength in Fig. \ref{fig:absorption}. As expected, larger losses are calculated at longer wavelengths because of the weaker mode confinement and the resulting overlap between the evanescent field and metal electrode. Given the total length of the actuated sections ($\approx 7$ mm) and the maximum calculated loss of 0.27 dB/cm at 420 nm, this would account for a reduction of the round-trip transmission ratio of about 0.98. Therefore, the metal electrodes contribute very little to the observed losses in the wavelength range considered here.  Given the negligible simulated loss contribution from the electrode, and the similar, but lower loss observed in \cite{Alumina_high_confinement, alumina_CMOS}, we are likely limited by sidewall scattering, and not absorption. Improved deposition and etching recipes should continue to reduce losses in future fabrication runs.

\begin{center} {\bf S2-E: Uncertainties in Intrinsic Losses} \end{center}
The accuracy that we can determine the FSR of the cavity ultimately determines the uncertainties in our reported linewidth and the intrinsic losses. We can estimate the error in our indirect approach at 320 nm by comparing the calculated FSRs at 369 nm and 420 nm with the directly measured FSRs acquired by scanning the ECDLs. Referring to Table \ref{tab:FSR}, we see a disagreement of $3.5\%$ and $5.7\%$ in the simulated FSRs compared to the measured values at 369 nm and 420 nm, respectively. This gave us confidence in the accuracy of the model. 

We also compared the fitted linewidths from the thermal scans at 369 and 420 nm with directly measured linewidths. We acquired linewidth measurements at 13 different chip temperatures at 369 nm and 10 different temperatures at 420 nm, each temperature bringing the chip into a different optical resonance. For each temperature, we tuned the ECDL wavelength across the resonance while monitoring the cavity transmission and the laser wavelength, measured with a commercial wavemeter. We found average linewidths of $6.14\pm0.09$ GHz (loaded Q of 130,000) at 369 nm and $4.78\pm0.16$ GHz (loaded Q of 140,000) at 420 nm. The errors here are the standard deviation of the 13 and 10 measured linwidths, respectively. These values compared well with the values found by fitting the thermally tuned data where we extracted loaded linewidths of 6.6 GHz at 369 nm (relative difference of 7.5\%) and 4.9 GHz at 420 nm (relative difference of 2.5\%). Since both the simulated FSR and fitted linewidth values are slightly overestimated when compared to directly measured values at 369 nm and 420 nm, we extrapolate that the loss and linewidth we reported at 320 nm are also overestimates, likely within 10\% of the true value. Therefore, we assume that the uncertainty in the calculated intrinsic losses is bounded by $\approx 10\% $.
\begin{center} {\bf \large S3: Simulations} \end{center}
Here we discuss the physics simulations used to estimate the tuning rates, refractive and group indices of the optical modes, and the FSRs. We used commercial multiphysics finite element software (COMSOL Multiphysics) to simulate the relevant physics and standard Python libraries to analyze the results.

\begin{center} {\bf S3-A: Material Properties} \end{center}
We used standard handbook values for the bulk mechanical densities and Young's moduli for silicon, silica, alumina, and aluminum. For aluminum nitride, we used the piezoelectric values listed in \cite{AlN_mat} multiplied by 0.9 based on the trend of the decreasing piezoelectric coefficients as a function of film thickness \cite{AlN_thick}, and the Young's modulus listed in \cite{AlN_mat2}. To model the optical waveguide modes and dispersion, we used the Sellmeier coefficients from \cite{SiO2_mat} for the silica cladding and Sellmeier coefficients derived from ellipsometry data for our alumina thin film. We used a one-term Selmier model to fit the ellipsometry  data (Eq. \ref{sellmeier}) and found $B=1.574$ and $C=1.581\times10^{-2}$ for $\lambda$ reported in microns.

\begin{equation}\label{sellmeier}
    n(\lambda) = 1 + B\frac{\lambda^2}{\lambda^2-C}
\end{equation}
We used interpolated extinction coefficients from \cite{Al_loss} to model the absorption loss from the top metal electrode.

To the best of our knowledge, the photoelastic coefficients of amorphous alumina thin films have not been reported in the literature. However, the photoelastic constants of the alumina component of composite aluminosilicate fibers are reported near 1550 nm in \cite{Al2O3_pe}. These values were in agreement with measurements done at 644 nm for crystalline aluminum oxide \cite{ruby_pe}. Therefore, we used the values from \cite{Al2O3_pe} to model the photoelastic behavior of our thin film at UV wavelengths.

\begin{center} {\bf S3-B: Free Spectral Range} \end{center}
The free spectral range in terms of frequency, $\nu_{FSR}$, of a general resonator with varying group index can be calculated with the formula
\begin{equation}\label{group_fsr}
    \nu_{FSR} = 1/t_{g} = c\left(\int ds \cdot n_g(s)\right)^{-1}
\end{equation}
where $t_g$ is the round trip group delay, $s$ parametrizes the path taken along the resonator, $n_g(s)$ is the position dependent group index, and $c$ is the speed of light. In order to evaluate this expression, we needed to model the effective group index for varying core widths since the resonator geometry adiabatically changes from a single-mode width (250 nm) for the curves to a multimode width (5000 nm) in the straights. We assumed a 150 nm thick waveguide core, sidewalls etched at $78$\unit{\degree}, and a top cladding thickness of 430 nm based on a cross sectioned SEM image. For each simulated width, we performed a numeric mode analysis and calculated the effective mode index at $\lambda$ and $\lambda\pm 1$ \unit{\nano\meter} for the vacuum wavelengths $\lambda=$ 320, 369, and 420 \unit{\nano\meter}. We calculated a numerical derivative using central differences (centered at $\lambda$) for each simulated width. Using these values, we calculated the position group index, Eq. \ref{group_n}.

\begin{equation}\label{group_n}
    n_g(\omega,s) = n_{eff}(\omega,s) + \omega\frac{\partial n_{eff}(\omega,s)}{\partial\omega} 
\end{equation}

Eq. \ref{group_fsr} was then numerically evaluated using Eq. \ref{group_n}. The final results are tabulated in Table \ref{tab:FSR}. Repeating the analysis and allowing the thickness of the alumina film to vary by 10\unit{\percent}, we found that the simulated FSRs vary by less than $2\%$. Given the direct relationship between the thickness of an ALD film and the number of deposition cycles, we believe that our targeted film thickness (150 \unit{\nano\meter}) is within $\approx1\%$ of the nominal value. Therefore, the disagreement between the simulated and measured values are likely due to error in the assumed optical indices of refraction. 

\begin{figure*}[ht]
    \centering
    \includegraphics[width=\textwidth]{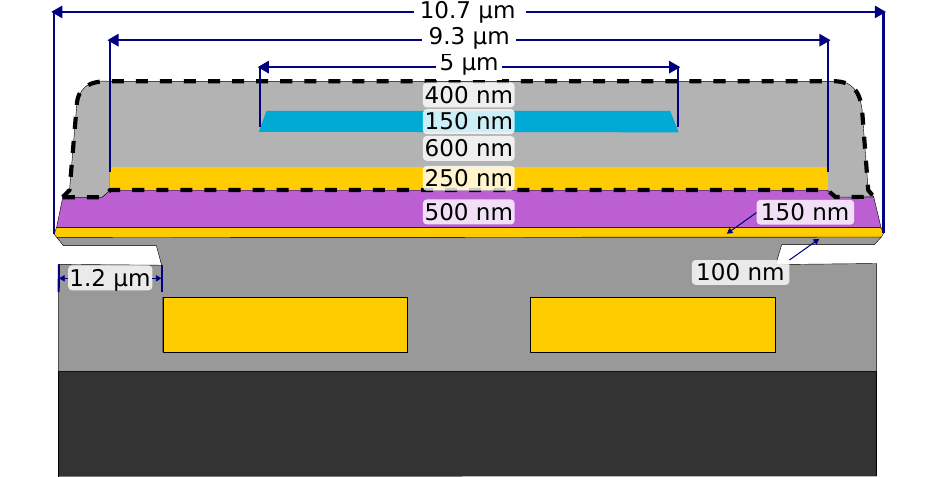}
    \caption{Cross section geometry used for strain tuning and electrode loss measurements. The region bordered by the dashed line indicates the geometry simulated for the electrode loss. An air domain surrounding the device was included for the optical mode simulations, but not shown here. The vertical layers are not to scale to improve readability.}
    \label{fig:cross_section_annotated}
\end{figure*}

\begin{figure*}[ht]
    \centering
    \includegraphics[width=\textwidth]{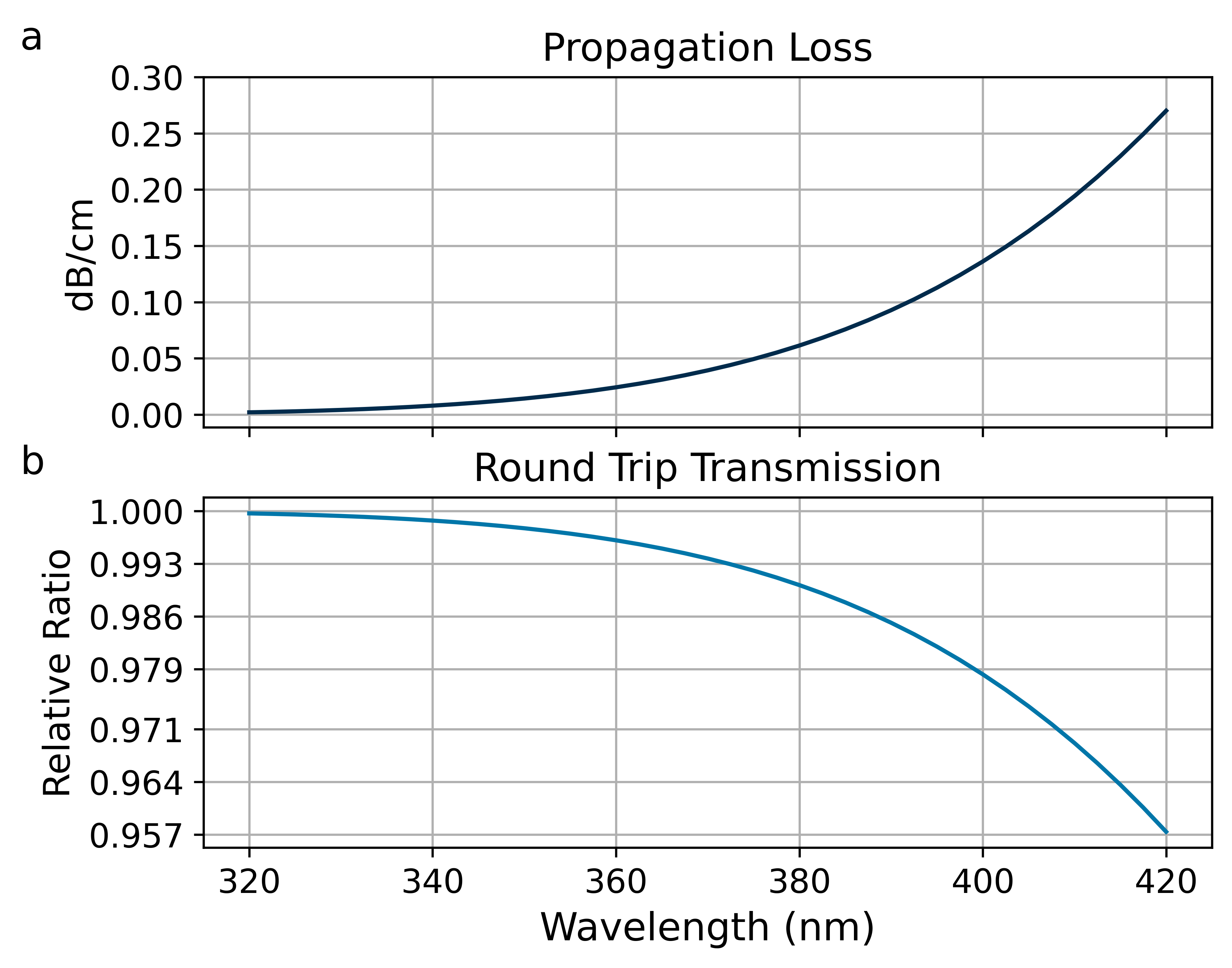}
    \caption{Simulated absorption loss from the top electrode. {\bf{a}} The propagation loss of the actuated portion of the resonator as a function of wavelength. {\bf{b}} The round trip transmission due to the actuators using $L\approx7$ \unit{mm}.}
    \label{fig:absorption}
\end{figure*}

\begin{center} {\bf S3-C: Optomechanical Coupling and Simulated Tuning Rates} \end{center}
In order to calculate the expected tuning rate per applied volt, it was necessary to construct a piezoelectric-mechanical coupled model and solve for the mechanical displacements per applied volt. To this end, we constructed a 2D simulation cross section (Fig. 1C. main text) with dimensions based on an SEM (Fig. 1B. main text). An annotated structure showing the layer thicknesses is given in \ref{fig:cross_section_annotated}. We also used an electromagnetic mode solver to calculate the fundamental TE mode of the cross section. We repeated the electromagnetic mode solver calculation, but with the displacements from the piezoelectric-mechanical study defining a moving mesh. The difference in the angular wavenumbers between the two mode studies is exactly the moving boundary perturbation, $\Delta\beta_{MB}$, up to the accuracy of the model.

To calculate the contribution from the photo-elastic effect, $\Delta\beta_{PE}$, we performed standard first-order perturbation calculations. With the piezoelectric-mechanical and mode study solutions, we calculated the photoelastic changes to the impermeability tensor $\eta_{ij}=1/\epsilon_{ij}$, where $\epsilon$ is the relative dielectric tensor \cite{photo_elastic}. Using the photoelastic relationship in strain-form, we have 

\begin{equation}\label{eq:PE}
    \Delta\left(\frac{1}{n^2}\right)_i =  p_{ij}S_j
\end{equation}
 where $\left(\frac{1}{n^2}\right)_i$ is a component of the perturbed impermeability tensor, $p_{ij}$ are components of the strain optic tensor and $S_j$ is a strain tensor component. Note that all tensors in Eq. \ref{eq:PE} are in contracted notation \cite{tensor_to_matrix}. From this expression it follows that
\begin{equation}
    \Delta n_i = -\frac{n^3}{2}p_{ij}S_j
\end{equation}
and
\begin{equation}
    \Delta \epsilon_i = -\epsilon_0 n^4 p_{ij}S_j
\end{equation}
where $n$ is the refractive index of the material, $\epsilon_{0}$ is the permittivity of free space, and $\Delta \epsilon_i$ is the change in the $i^{th}$ component of the dielectric permittivity tensor. Finally, we evaluated the standard expression \cite{perturbation} for perturbations to the angular wavenumber of a waveguide,
\begin{equation}
    \Delta\beta_{PE} = \omega \frac{\int_{S}dS \left(E_i^*\Delta\epsilon_{ij}E_j\right)}{\int_S d\vec{S}\cdot (\vec{E}^*\times\vec{H}+\vec{E}\times\vec{H}^*)}
\end{equation}
where $\omega$ is the unperturbed angular frequency of the mode, the integrals are taken over the cross sectional area of the waveguide $S$, $E_i (H_i)$ is a component of the electric (magnetic) field vector $\vec{E} (\vec{H)}$, and $\Delta\epsilon_{ij}$ has been converted from contracted notation back to Cartesian indices.

The simulated results are tabulated in Table \ref{tab:perturbation}. As mentioned in the main text, the simulated tuning rates are several times larger than what we observed. We believe that this discrepancy is due to differences in the assumed piezoelectric and photoelastic properties of our particular films. Even though the grain orientation of our AlN film is highly aligned, we believe that the piezoelectric constants are somewhat reduced compared to the literature, and that the grain orientation can be improved in future fabrication runs. We also assumed photoelastic constants based on different forms of $\text{Al}_2\text{O}_3$, and not values based on thin films because of the lack of published values.  Additionally, the wavelengths used to measure the photoelastic constants in \cite{Al2O3_pe,ruby_pe} were not at near-UV and UV wavelengths. Early photoelastic studies of amorphous silica showed dispersion in the stress-optic coefficients at low wavelengths \cite{PE_dispersion_SiO2}, so our assumed values used in the simulations may not be correct at these wavelengths.

\begin{figure*}[ht]
    \centering
    \includegraphics[width=\textwidth]{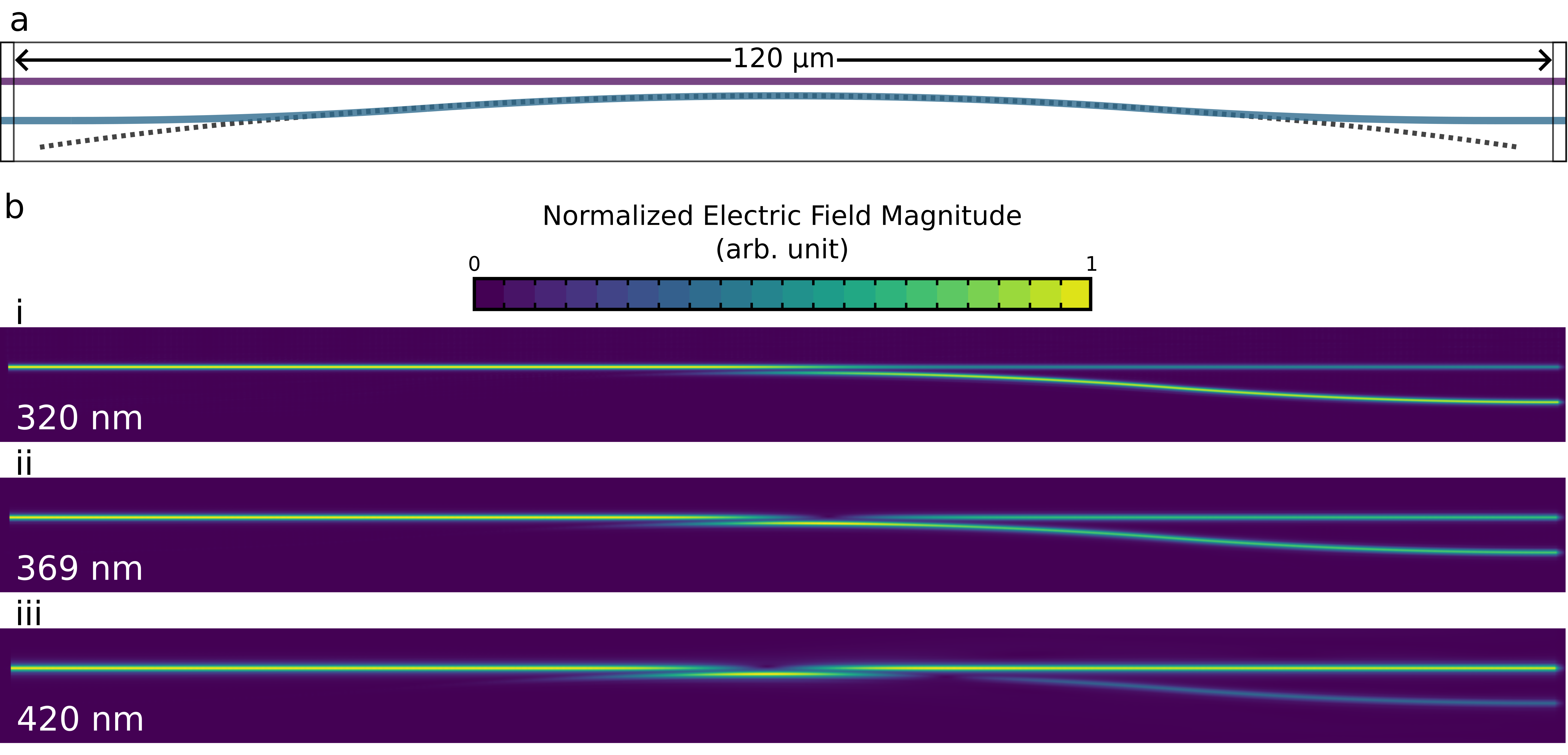}
    \caption{Bus-resonator coupling simulations.  {\bf{a}} Simulation bounds are shown as solid black lines. The bus waveguide is purple and the approximated resonator waveguide is blue. The uninterrupted arc of the resonator is shown as a dashed curve. A constant waveguide thickness is assumed, with rectangular core cross sections that are 250 nm wide. {\bf{b}} Evolution of the TE polarized electric field mode when launched from the left hand side of the bus waveguide.}
    \label{fig:coupling_3D}
\end{figure*}

\begin{center} {\bf S3-D: Bus-Resonator Coupling} \end{center}
As a starting point to simulate the cross-coupling coefficient, $t_1$, at each wavelength, we constructed a 3D model geometry and assumed waveguide widths and a coupling gap based on the drawn lithography mask features. We also assumed vertical side walls. A cross section of the simulation domain, centered on the waveguide core thickness, is shown schematically in Fig. \ref{fig:coupling_3D} a. Both the bus and resonator arc are assumed to have rectangular cores, 250 nm wide and 150 nm thick. The coupling was drawn at 205 nm at the apex of the resonator. We approximate the full curve by taking an 8.6 degree arc with a 400 \unit{\micro\meter} radius of curvature, which then curves back so that the input faces are parallel with the bus waveguide at the end of the simulation domains. The cores are separated by at least 10 core widths at the ends of the simulation domains. The separation exceeds 5.4 core widths when the lower waveguide has the correct curvature and begins to approximate the resonator arc. COMSOL's beam envelope method was used to calculate the field evolution in the simulation volume, with results shown in Fig. \ref{fig:coupling_3D} b. The simulated cross-coupling coefficient, $t_1$, is readily extracted by taking the square root of the power transmittance observed in the lower right hand port. We found electric field cross-coupling coefficients of approximately 0.9, 0.7, and 0.8 at 320, 369, and 420 nm, respectively. Although there appears to be good agreement at 320 nm when compared to $t''$, the other values exceed the fitted parameters by at least $17\%$. Taking this at face value would predict a peculiar trend in the loss values. 

\begin{figure*}[ht]
    \centering
    \includegraphics[width=\textwidth]{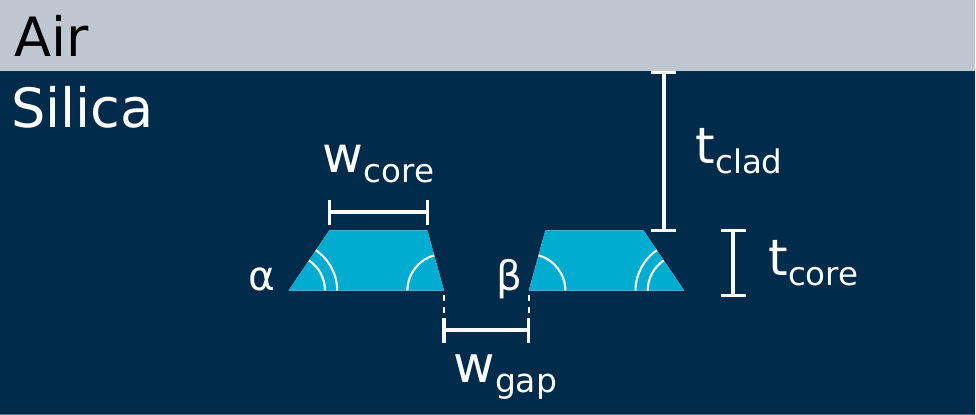}
    \caption{Cross section geometry used for super-mode simulations. The blue trapezoids are the alumina waveguides with exterior sidewall angle of $\alpha=75^{\circ}$ and interior sidewall angle $\beta=48^{\circ}$. Critical dimensions are annotated. Values used in 2D simulations: $\text{w}_{\text{core}}=300 \text{ nm}$, $\text{t}_{\text{clad}}=430 \text{ nm}$, $\text{t}_{\text{core}}=150 \text{ nm}$. The gap between the waveguides, $\text{w}_\text{gap}$, is varied between 150 and 250 nm.
    }
    \label{fig:coupler_sketch}
\end{figure*}

\begin{figure*}[ht]
    \centering
    \includegraphics[width=\textwidth]{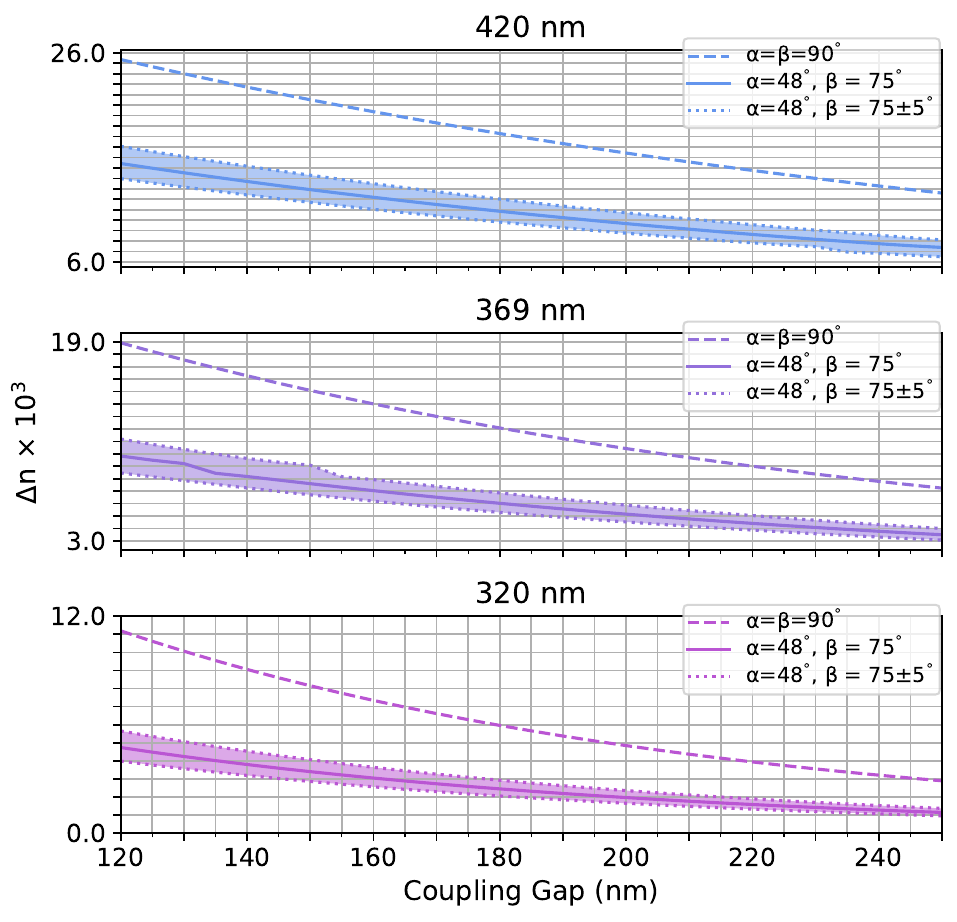}
    \caption{Simulated effective index contrast, $\Delta n$, between the first two TE polarized super-modes, using the geometry detailed in Fig. S7. The solid lines were calculated using our inferred sidewall angles near the apex of the resonator. The difference between the dashed/dotted lines with the solid lines illustrate the significance that the sidewall angles have on the index contrast, and hence the coupling between the waveguides.
    }
    \label{fig:coupler_sim}
\end{figure*}

To better understand the discrepancy in the simulated and potential cross-couplings, we examined critical dimension scanning electron micrographs (CD-SEM) of comparable structures before the top oxide was deposited.  Because the CD-SEMs were taken before the top oxide layer was deposited, similar images can not be taken for the device tested.
The CD-SEMs indicate that the drawn gap  of 205 nm was likely as small as 120 nm at the bottom of the film. Furthermore, based on cross-sectioned SEMs, the assumption that the sidewalls are vertical is not correct. Based on the CD-SEMs' image contrasts near the waveguide edges, we estimated that the exterior sidewall angles ($\alpha$ in Fig. \ref{fig:coupler_sim}) are approximately 48\unit{\degree} and that the interior angles ($\beta$ in Fig. \ref{fig:coupler_sketch}) are approximately 75\unit{\degree} at the designed waveguide spacing of 205 nm. The interior sidewall angle also varied with the coupling gap, approaching the exterior angle far from the resonator apex. These features motivated simpler 2D simulations to investigate the coupler's sensitivity to varying gaps and varying sidewall profiles.

\begin{center} {\bf S3-E: Coupling Sensitivity} \end{center}
To investigate the coupling sensitivity to varying fabrication features, we simulated the difference between the effective mode indices, $\Delta n_{eff}$, of the first two TE-polarized super-modes of the geometry shown in Fig. \ref{fig:coupler_sketch}. We varied the coupling gaps and sidewall angles, with results plotted in Fig \ref{fig:coupler_sim}. For a given coupling gap, $\Delta n_{eff}$,is significantly larger for vertical sidewalls when compared with angled sidewalls more consistent with the CD-SEMs, putting the accuracy of the initial 3D simulation into serious question. We also observed that small deviations of just 5\unit{\degree} about the assumed interior sidewall angle of 75\unit{\degree} cause changes around 10\unit{\percent} in $\Delta\beta_{TE}$. To get a basic sense of the significance of these variations, we assumed that the cross-coupling of the extended bus-resonator could be approximated by an effective length, $L_{eff}$, and an effective wavenumber difference, $\Delta\beta$. To wit,
\begin{equation}\label{eq:estimate}
    t_1 = \sqrt{\frac{P_2}{P_1}}=
    \left|
    \sin
    \left(
    \frac{L_{\text{eff}}\Delta\beta_{eff}}{2}\right)
    \right|
    =
        \left|
    \sin
    \left(
    \frac{\pi L_{\text{eff}}\Delta n_{eff}}{\lambda_0}\right)
    \right|
\end{equation}
Given that we want to predict values of $t_1\approx0.6$ accurately, we solved Eq. \ref{eq:estimate}, and then varied the argument of the right hand side by 10\unit{\percent}. The calculated $t_1$ is as small as 0.55 and as large as 0.86. This illustrates the need for a very high degree of accuracy in the modeled geometry. However, inferring the geometry of the bus and resonator waveguide along the coupling region is nontrivial. Because of the larger radius of curvature of the resonator arc (400 \unit{\micro\meter}), the effective coupling length is 10's of \unit{\micro\meter}. Accurately determining the exact waveguide geometries over this length via cross-sectioned SEMs is therefore highly impracticable.

\pagebreak

\pagebreak
\clearpage

\clearpage

\newpage

\begin{table}[H]    
    \centering
    \begin{tabular}{|c|c|c|c|c| }

        \hline
         Wavelength (nm) & t=135 nm & t =150 nm & t=165 nm & Measured Value \\
         \hline
        320 & 5.48 pm & 5.58 pm & 5.54 pm & N/A \\
        \hline
        369 & 7.73 pm & 7.75 pm & 7.61 pm & $7.49\pm0.08$ pm \\
        \hline
        420 & 10.41 pm & 10.39 pm & 10.25 pm & $9.83\pm0.14$ pm \\
        \hline    
    \end{tabular}    
    \caption{Simulated and Measured Free Spectral Ranges. 
    The waveguide core thickness, t, was varied by 10\unit{\percent} from the targeted deposition thickness of 150 \unit{\nano\meter} for the simulations, showing minimal difference in the resulting FSRs for a fixed wavelength. Although not directly measured, we expect that our film thickness was 150 nm within $\approx1\%$ accuracy because of the direct relationship between the number of deposition cycles and the ALD film thickness.}
    \label{tab:FSR}
\end{table}

\begin{table}[H]
    \centering    
    \begin{tabular}{|c|c|c|c|c|c|c|}
        \hline
        Wavelength (nm) & $t'$  & $t''$ & $\alpha_{\text{dB/cm},1}$ & $\alpha_{\text{dB/cm},2}$& $Q_1$& $Q_2$ \\
        \hline
        320 & 0.594 & 0.847 & $4.4$& $1.4$&$350,000$&$1,100,000$\\
         \hline
        369 & 0.302& 0.603 &  $10$&  $4.3$&  130,000&  300,000\\
         \hline
        420 & 0.536 & 0.681&  $5.3$&  $3.3$& 200,000 &  330,000\\
        \hline
    \end{tabular}
    \caption{Extracted Model Values and Intrinsic Losses. We refer to fitted parameters as $t'$ and $t''$ because of the ambiguity in the coupling condition. A $10\%$ error is assume for the propagation losses and intrinsic quality factors.} 
    \label{tab:model_fits}
\end{table}

\begin{table}[H]
    \centering    
    \begin{tabular}{|c|c|c|c|}
        \hline
        Wavelength (nm) & $\Delta\beta_{PE}/V$ (rad/Vm) & $\Delta\beta_{MB}/V$ (rad/Vm)&$\int ds\beta(s)$ \\
        \hline
        320 & -15.9 & -1.05& 317,000\\
         \hline
        369 & -10.3 & -0.936&  270,000\\
         \hline
        420 & -7.06 & -0.822& 234,000\\
        \hline
    \end{tabular}
    \caption{Simulated Perturbations to the angular Wavenumber, $\beta$ for a 1 V potential difference across the piezoelectric actuator. The total actuated length is $2L=6.9$ \unit{mm}. The last column is the simulated round trip phase accumulation, corresponding to the denominator in Eq.3 of the main text. }
    \label{tab:perturbation}
\end{table}
\pagebreak
\clearpage

\bibliographystyle{IEEEtran}
\bibliography{main_supplement}